\documentclass[prd,superscriptaddress,nofootinbib,amsmath,amssymb,aps,11pt]{revtex4-2}

\usepackage{bm}
\usepackage{amsfonts}
\usepackage{latexsym}
\usepackage[latin1]{inputenc}
\usepackage{graphicx}
\usepackage{amsmath}
\usepackage{tabularx}
\usepackage{array}
\usepackage{dcolumn} 
\usepackage{palatino}
\usepackage{mathpazo}
\usepackage{appendix}
\usepackage{booktabs}
\usepackage{dcolumn}
\usepackage[dvipsnames]{xcolor}

\def\jnl@style{\it}
\def\aaref@jnl#1{{\jnl@style#1}}

\def\aaref@jnl#1{{\jnl@style#1}}

\def\aj{\aaref@jnl{AJ}}                   
\def\apj{\aaref@jnl{ApJ}}                 
\def\apjl{\aaref@jnl{ApJ}}                
\def\apjs{\aaref@jnl{ApJS}}               
\def\apss{\aaref@jnl{Ap\&SS}}             
\def\aap{\aaref@jnl{A\&A}}                
\def\aapr{\aaref@jnl{A\&A~Rev.}}          
\def\aaps{\aaref@jnl{A\&AS}}              
\def\mnras{\aaref@jnl{Mon.~Not.~Roy.~Astron.~Soc.}}             
\def\prd{\aaref@jnl{Phys.~Rev.~D}}        
\def\prc{\aaref@jnl{Phys.~Rev.~C}}  
\def\prl{\aaref@jnl{Phys.~Rev.~Lett.}}    
\def\qjras{\aaref@jnl{QJRAS}}             
\def\skytel{\aaref@jnl{S\&T}}             
\def\ssr{\aaref@jnl{Space~Sci.~Rev.}}     
\def\zap{\aaref@jnl{ZAp}}                 
\def\nat{\aaref@jnl{Nature}}              
\def\aplett{\aaref@jnl{Astrophys.~Lett.}} 
\def\apspr{\aaref@jnl{Astrophys.~Space~Phys.~Res.}} 
\def\physrep{\aaref@jnl{Phys.~Rep.}}      
\def\physscr{\aaref@jnl{Phys.~Scr}}       
\def\commat{\aaref@jnl{Comm.~Math.~Phys.}}              
\def\science{\aaref@jnl{Science}}               
\def\cqg{\aaref@jnl{Classical Quant.~Grav.}}            
\def\jpcs{\aaref@jnl{JPCS}}                                     
\def\ijmpd{\aaref@jnl{Int.~J.~Mod.~Phys.~D}}                    
\def\grg{\aaref@jnl{Gen.~Relat.~Gravit.}}               
\def\rpp{\aaref@jnl{Rep.~Prog.~Phys.}}          
\def\npa{\aaref@jnl{Nucl.~Phys.~A}}        
\def\lrr{\aaref@jnl{Living Rev.~Rel.}}                   
\def\jcap{\aaref@jnl{J.~Cosmology Astropart.~Phys.}}    
\def\rmp{\aaref@jnl{Rev.~Mod.~Phys.}}   

\newcommand{\be}{\begin{equation}}
\newcommand{\ee}{\end{equation}}
 
\begin{document}	
	
\title{Testing disformal non-circular deformation of Kerr black holes with LISA}

\author{Eugeny Babichev}
\email{eugeny.babichev@ijclab.in2p3.fr}
\affiliation{Universit\'e Paris-Saclay, CNRS/IN2P3, IJCLab, 91405 Orsay, France}

\author{Christos Charmousis}
\email{christos.charmousis@ijclab.in2p3.fr}
\affiliation{Universit\'e Paris-Saclay, CNRS/IN2P3, IJCLab, 91405 Orsay, France}

\author{Daniela D. Doneva}
\email{daniela.doneva@uni-tuebingen.de}
\affiliation{Theoretical Astrophysics, Eberhard Karls University of T\"ubingen, T\"ubingen 72076, Germany}
\affiliation{INRNE - Bulgarian Academy of Sciences, 1784  Sofia, Bulgaria}
	
\author{Galin N. Gyulchev}
\email{gyulchev@phys.uni-sofia.bg}
\affiliation{Department of Theoretical Physics, Faculty of Physics, Sofia University, Sofia 1164, Bulgaria}

\author{Stoytcho S. Yazadjiev}
\email{yazad@phys.uni-sofia.bg}
\affiliation{Department of Theoretical Physics, Faculty of Physics, Sofia University, Sofia 1164, Bulgaria}
\affiliation{Institute of Mathematics and Informatics, 	Bulgarian Academy of Sciences, 	Acad. G. Bonchev St. 8, Sofia 1113, Bulgaria}
	

\begin{abstract}
There is strong observational evidence that almost every large galaxy has a supermassive black hole at its center. It is of fundamental importance to know whether such black holes are described by the standard Kerr solution in General Relativity (GR) or by another black hole solution. An interesting alternative is the so-called disformal Kerr black holes which exist within the framework of degenerate higher-order scalar-tensor (DHOST) theories of gravity. The departure from the standard Kerr black hole spacetime is parametrized by a parameter $D$, called  {\it disformal parameter}. In the present work, we discuss the capability of LISA to detect the disformal parameter. For this purpose, we study Extreme Mass Ratio Inspirals (EMRI's) around disformal Kerr black holes within the framework of the quadrupole hybrid formalism. Even when the disformal parameter is very small, its effect on the globally accumulated phase of the gravitational waveform of an EMRI  can be significant due to the large number of cycles in the LISA band made by the small compact object. We show that LISA will in principle be able to detect and measure extremely small values of the disformal parameter which in turn,  can be seen as an assessment of LISA's ability to detect very small deviations from the  Kerr geometry. 
\end{abstract}
	
    \maketitle
	
\section{Introduction}
Gravitational wave (GW) astronomy is a powerful tool for fundamental studies in physics \cite{Barack:2018yly,Carson:2020rea}. It opens a new window toward testing the strong field regime of gravity and even fundamental symmetries of spacetime. The remarkable success of the ground-based gravitational detectors gives us confidence in the great importance of the future space-based ones that are designed to detect GW signals of much lower frequencies. Such a space-based detector, which is at an advanced stage, is the Laser Interferometer Space Antenna (LISA). It is expected to open windows to various astrophysical phenomena so far inaccessible to us, among which extreme mass ratio inspirals (EMRIs) \cite{LISA:2022kgy, LISA:2022yao}. 

EMRIs constitute a small compact object (SCO), (e.g. a stellar-mass black hole or a neutron star) spiraling into a supermassive black hole (SMBH) in a galactic center with masses of the order of $10^5 - 10^7 M_{\odot}$.  The large difference of mass scales, typically $10^{-6} - 10^{-4}$, means that the SCO influence on the binary system may safely be considered to be a perturbation of the SMBH spacetime. The SCO is expected to undergo typically about $10^{5} - 10^{6}$ cycles around the SMBH in the LISA frequency band before plunging. 
This fact is  important for fundamental physics because even very small deviations caused by possibly new physics accumulate and eventually could be detected by LISA and other future space-based detectors. For example, EMRIs can probe interesting features such as the presence of a scalar field around the primary black hole \cite{Collodel:2021jwi,Delgado:2023wnj,Brito:2023pyl} or even the secondary one \cite{Maselli:2021men}, the presence of dark matter \cite{Rahman:2022fay}, nonintegrable deformations \cite{Destounis:2021mqv,AbhishekChowdhuri:2023wdf},  parity violation \cite{Sopuerta:2009iy,Pani:2011xj,Canizares:2012is}, etc. \cite{Rahman:2023sof,Zi:2023qfk,LISAConsortiumWaveformWorkingGroup:2023arg}. 

There is strong observational evidence that almost every large galaxy has a SMBH at its center. For example, our galaxy hosts a SMBH corresponding to the radio source Sagittarius A*. A fundamental question that needs to be answered is whether the central objects are described by the standard Kerr solution in General Relativity (GR) or by another black hole solution. Even though until now some more exotic geometries were ruled out mainly by the Event Horizon observations \cite{Volkel:2020xlc, Kocherlakota}, their accuracy is still not good enough to have a conclusive answer. LISA will be able to fill this gap \cite{LISA:2022kgy,LISA:2022yao}.

Certain modified gravity theories predict the existence of black holes (BHs)  different from Kerr \cite{Herdeiro:2015waa, Babichev:2023psy}. According to the current gravitational wave observations \cite{LIGOScientific:2020tif,LIGOScientific:2021sio,Wong:2022wni,Lyu:2022gdr}, if such new type BHs exist, the astrophysical deviations from GR should be small enough. Therefore, in order to test the possible existence of a novel BH the observational tools should be sensitive enough. The EMRIs observed in the future by LISA are supposed to be able to serve this purpose. With this motivation in mind, in the present paper, we study EMRIs around a new type of BH dubbed {\it disformal Kerr BHs} which were constructed within the framework of degenerate higher-order scalar-tensor (DHOST) theories of gravity \cite{Anson:2020trg}. 
The {\it disformal parameter} $D$, measuring roughly speaking deviation from GR, 
enters the disformed metric and  labels a 
DHOST theory admitting the disformed metric as a solution (for the relevant DHOST formulas see \cite{BenAchour:2020fgy}). In this sense, the disformal parameter $D$ does not correspond to extra hair (because changing $D$ changes the theory), but rather it identifies a new black hole solution in a specific DHOST theory.

The disformal Kerr metric is one of the very few stationary solutions that provides a measurable departure from Kerr. 
Being an explicit and simple enough solution, its properties can be studied in parallel with Kerr in a more or less straightforward fashion.
Like all explicit solutions it has a number of positive but also negative points which stem from its explicit form and construction properties.
Disformal Kerr is constructed from the stealth Kerr solution~\cite{Charmousis:2019vnf} which is Kerr spacetime accompanied by a non trivial scalar field with a radial profile and a linear time dependence. It turns out~\cite{Charmousis:2019vnf}, that the  scalar field is associated to a geodesic congruence of Kerr. The scalar is in fact a particular Hamilton-Jacobi functional depicting integrable Kerr geodesics which is chosen to be regular. As such in stealth Kerr the scalar hair is very particular and accommodates spacetime in a geodesic and regular fashion. The tensor perturbations obey a modified Teukolsky equation~\cite{Charmousis:2019fre} but the scalar perturbations are frozen and the scalar does not obey a hyperbolic type of equation~\cite{Babichev:2018uiw,deRham:2019gha}. 
This most probably renders the scalar perturbations of the disformal Kerr solution pathological. On the positive side the metric is surprisingly regular, does not contain closed causal curves, and has other non trivial 
properties. This is thanks to the fact that the deformation is associated to geodesics of the Kerr metric itself~\cite{Charmousis:2019vnf}. Given the difficulty of obtaining explicit stationary spacetimes, ad hoc metrics are often discussed in the literature (see for example \cite{Johannsen:2013szh} and more recently \cite{Heumann:2022cyi} and references within). However, ad hoc metrics often do not share regularity and/or causality of the dismorfed Kerr metric being a consequence of the nature of their ad hoc construction.

Disformal Kerr has many interesting properties which differentiate it from the Kerr metric \cite{Anson:2020trg}: the event horizon no longer lies at constant
radius $r$ and is not a Killing horizon; the limiting surface for stationary observers is generically distinct from the outer event horizon; the metric is non-circular, meaning that it cannot be written in a form that exhibits the reflection symmetry $(t,\varphi)\to (-t, -\varphi)$. The latter property is particularly interesting because such symmetry is usually assumed for axisymmetric spacetimes in the literature as it is a  property of Ricci flat spacetimes in GR. The well defined scalar field also guarantees spacetime to be stably causal. Therefore, the disformal Kerr metric gives an interesting counter solution to Kerr and the current work can be considered also as a model study which leads to a better understanding of the implications of non-circularity and the differing properties discussed above. Also the metrics at hand are one of the not-so-many interesting solutions of potentially super-massive non-Kerr black holes.

In this paper we will take a conservative viewpoint and consider $D$ as a deformation parameter for a whole class of metrics rather than treat each $D$ deformation as a DHOST solution. 
In other words, we consider the disformal Kerr solution as a metric deformation of the Kerr spacetime. 
Moreover, when applied our analysis of EMRIs to the gravitational radiation, we will calculate gravitational radiation assuming standard GR analysis. Such an approach contains a very important assumption, namely, we take into account the metric deformation of the GR solution (in our case Kerr deformation), while keeping the perturbations to be of the standard GR form.

The paper is organized as follows. In section \ref{DBHS} we briefly discuss the disformal black hole solution and some of its main characteristics. Section \ref{EMRIS} is devoted to a general discussion of the circular geodesics on the equatorial plane and the EMRIs within the framework of the quadrupole hybrid formalism. Our numerical results for the prograde and retrograde EMRIs are presented in section \ref{NR} and the Apandix. In the last section \ref{Detection} we estimate the minimal possible disformal parameter that could in principle be detected by LISA. The paper ends with a Discussion.

\section{Disformal black hole solution}\label{DBHS}

The explicit form of the disformal Kerr metric in Boyer-Lindquist type coordinates is given by~\cite{Anson:2020trg}
\begin{equation}
	\begin{split}
		\label{df}
		\tilde{g}_{\mu\nu}d x^\mu d x^\nu &= -\left(1-\frac{2Mr}{\rho^2}\right)\ dt^2 -\frac{4Mar\sin^2\theta}{\rho^2} dtd\varphi + \frac{\sin^2\theta}{\rho^2}\left[\left(r^2+a^2_*\right)^2-a_*^2\Delta
		\sin^2\theta\right] d\varphi^2\\
		&+  \frac{\rho^2 \Delta - 2 M(1+D) r D (a^2_*+r^2)}{\Delta^2} d r^2 - 2D \frac{\sqrt{2Mr(a^2_*+r^2)}}{\Delta} d td r + \rho^2 d \theta^2\; .
	\end{split}
\end{equation}
Here $\Delta$ and $\rho$ are defined as
\begin{eqnarray}
\Delta= r^2 + a^2_*- 2M_*r, \, \,\, \, 
\rho= r^2 + a^2_* \cos^2\theta,
\end{eqnarray}
with $M_*$ and $a_*$ being
\begin{eqnarray}
M_*= (1+D)M , \, \,\, \, 
a_* = \frac{a}{\sqrt{1+ D}}.
\end{eqnarray}
The parameters $M$ and $a$ are the physical mass and the angular momentum per unit mass as seen at spacial infinity, while $D>-1$ is the dimensionless disformal parameter. In fact $M_*, a_*$ are the mass and angular momentum of the seed stealth Kerr metric, however, in the context of this paper they can be viewed as auxilary parameters. As we will see expicitely in a moment when $a=0$ we get back to a Schwarzschild geometry we have thus an identical static limit. Otherwise if $a\neq 0$ it is only when $D=0$ that we have a Ricci flat metric. It is therefore clear that the disformed Kerr BH is a deformation of the Kerr spacetime. Note also that the metric shares an ergosphere and the same ring singularity at $\rho=0$ as Kerr. The spacetime is also stably causal as the scalar field itself as its gradient gives a future directed timelike vector field \cite{Anson:2020trg}.

A non-trivial property of the new metric, that distinguishes it from the Kerr metric,  is the term $g_{tr}$, which cannot be eliminated by a coordinate change without introducing other off-diagonal elements. On the other hand, in the static limit case $a=0$, the off-diagonal term $g_{tr}$ in~(\ref{df}) can be removed by the following coordinate transformation:
\begin{equation}
	d t = d T -
	\frac{D\sqrt{2Mr^3}}{\Delta\left(1-\frac{2M}{r}\right)}d
	r\; ,
\end{equation}
Thus the resulting metric is nothing but the Schwarzschild spacetime:
\begin{equation*}
	\tilde{g}_{\mu\nu}d x^\mu d x^\nu =
	-\left(1-\frac{2M}{r}\right) d T^2 +
	\left(1-\frac{2M}{r}\right)^{-1} d r^2 + r^2 d \Omega^2\;
	.\end{equation*}

In studying the EMRIs we shall consider orbits on the equatorial plane $\theta=\frac{\pi}{2}$ only and that is why we will need the restriction of the spacetime metric on that plane. It turns out convenient to trade the off-diagonal term $g_{tr}$ for $g_{r\varphi}$ by the coordinate transformation    
\begin{equation*}
   		d t \to d t  -
   	D\frac{\sqrt{2Mr(r^2 + a^2_*)}}{\Delta\left(1-\frac{2M}{r}\right)} d r . 
\end{equation*}  

Then the restriction of the spacetime metric on the equatorial plane $\theta=\frac{\pi}{2}$ reads
\begin{equation}
	\begin{split}
		\label{df1}
		 g_{\mu\nu} d x^\mu d x^\nu &= -\left(1-\frac{2M}{r}\right) d t^2 -\frac{4Ma}{r} d t d\varphi + \left[r^2+a^2_* + \frac{2M_*a^2_*}{r}
		\right] d\varphi^2\\
		& + \frac{1}{\Delta^2}\left[r^2 \Delta - \frac{1-\frac{2M_*}{r}}{ 1- \frac{2M}{r}} 2MDr (r^2 + a^2_*) \right] d r^2 
		+ 4D \frac{Ma}{r} \frac{\sqrt{2Mr(a^2_*+r^2)}}{\Delta(1- \frac{2M}{r})} d r d \varphi .
	\end{split}
\end{equation}
In the appendix we give a coordinate system adapted to the equator which is circular. We nevertheless choose to work with the above coordinate system for the sake of generality and for future reference. 

When the horizon exists for given parameters its cross section with  the equatorial plane is given by the equation
\begin{equation}
\label{statlimit}
	\Delta_{eq}=r^2 + \frac{a^2}{D+1} - 2Mr + 2D\frac{Ma^2}{r (D+1)}=0.
\end{equation} 
The conditions for existence of the horizon are, however, stronger than those of Eq.~(\ref{statlimit}). Following~\cite{Anson:2020trg} one can separate two cases, depending on the sign of $D$. In the case $D<0$, the condition for the existence of the horizon is $a_*<a_{*c}$, where $a_{*c}$ is the critical value of $a_*$, given by the solution of the fourth order polynomial (see Appendix B of~\cite{Anson:2020trg}):
\begin{gather*}
-256 D^2 + 32 D\left[39 + D\left(50-13D\right)\right]\frac{{a_{*c}^2}}{M^2} + \left[15 + D\left(343 D^3 + 2324 D^2 + 562 D -2076\right)\right]\frac{a_{*c}^4}{M^4}\\ - 2 \left[15- D\left(414 + 517 D\right)\right]\frac{a_{*c}^6}{M^6} + 15 \frac{{a_{*c}^8}}{M^8} =0.
\end{gather*}
Note that $\frac{a_*}{M}<1$ as it is shown in Fig.~1 of \cite{Anson:2020trg}. It is not difficult to see that in this case Eq.~(\ref{statlimit}) has always a solution. 
In the case $D>0$ there is an analogous, but simpler condition for the horizon existence, namely $a_*<a_{*c}$, where now
$$
\frac{a_{*c}}{M} = \frac{1}{\sqrt{1+4D}} \quad \Rightarrow \quad \frac{a}{M}< \sqrt{\frac{1+D}{1+4D}}.
$$
Again, this condition is stronger than the bound on the possible values of $a_*$ and $D$ from Eq.~(\ref{statlimit}).

\begin{figure}
	\includegraphics[width=0.8\textwidth]{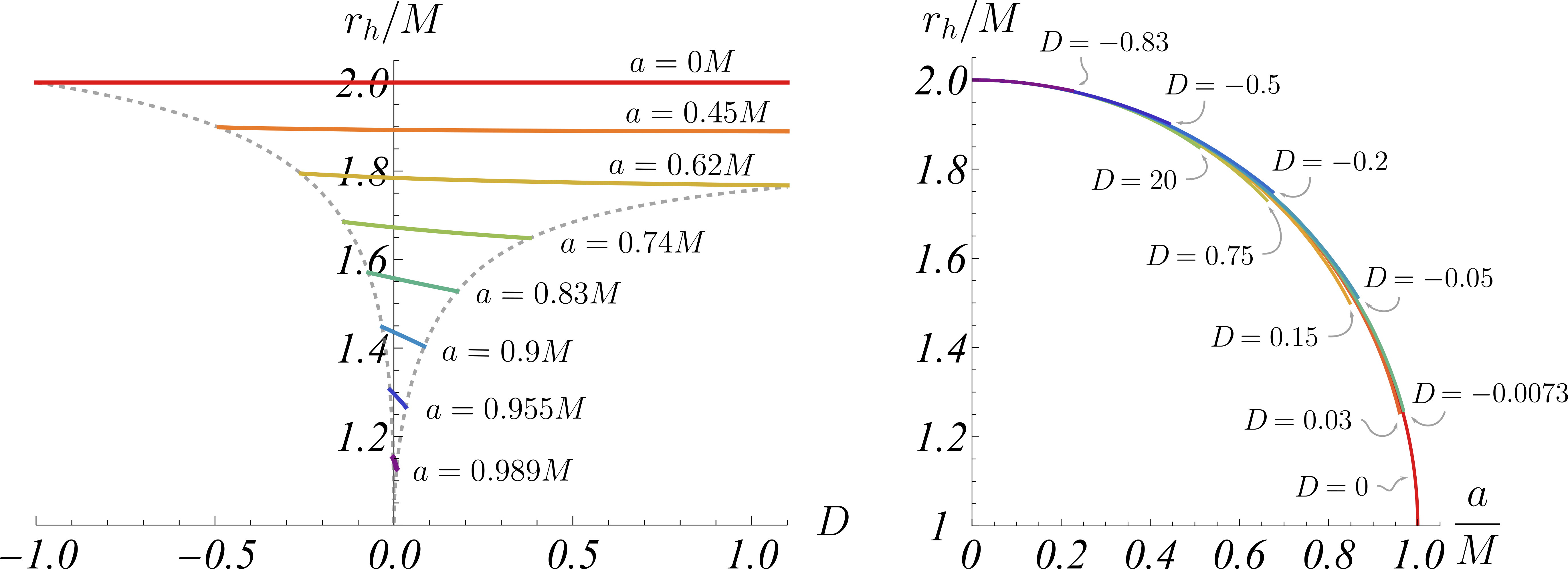}
	\caption{The radius of the black hole horizon $r_{h}/M$ denoted with solid lines as a function of the disformal parameter $D$ for different values of the black hole spin parameter $a/M$ (\textit{left panel}) and as a function of $a/M$ for different values of the disformal parameter (\textit{right panel}). The gray dashed lines on the left panel correspond to those values of the disformal and spin parameters beyond which the black hole horizon doesn't exist.}
	\label{fig:r_Horizons} 
\end{figure}

In Fig. \ref{fig:r_Horizons}, we show the dependence of the black hole horizon on the disformal parameter $D$ for different values of the BH spin parameter $a/M$. It is seen that with the increase of the disformal parameter as well as the BH angular momentum, the horizon radius decreases.

It should also be noted that the disformed Kerr solution (\ref{df}) has a well defined limit for $D\to \infty$ keeping the physical spin 
parameter $a$ fixed \cite{Anson:2021yli}.  The restriction of that limit  on the equatorial plane can be easily obtained from (\ref{df1}) 
in  the limit $D\to \infty$ for fixed $a$, namely 

\begin{equation}
		\label{limit_df1}
		 ds^2 = -\left(1-\frac{2M}{r}\right) d t^2 -\frac{4Ma}{r} d t d\varphi + 
   \left(r^2  + \frac{2M a^2}{r}
		\right) d\varphi^2
		+  \frac{d r^2}{1- \frac{2M}{r}} 
		- \frac{2a}{r} \frac{\sqrt{2Mr}}{1-\frac{2M}{r}}  d r d \varphi .
\end{equation}

\section{Extreme mass ratio inspirals}\label{EMRIS}

\subsection{Circular geodesics}

As discussed, we shall consider orbits restricted to motion on the equatorial plane $\theta=\frac{\pi}{2}$. There are two integrals of motion for an orbiting particle (or a SCO) around the black hole, namely the energy $\frac{E}{m}=-u_t$ and the angular momentum $\frac{L}{m}=u_\varphi$, respectively, where $m$ is the mass of the orbiting object. The equation of motion in the radial direction is taken from the norm of the normalized four-velocity $u^{\mu}u_{\mu}=-1$,
\begin{equation}
	\label{eq:rd}
	{\hat g}_{rr}\left(\frac{dr}{d\tau}\right)^2=V(r)
\end{equation}
where we defined an effective potential 
\begin{equation}
	\label{vpm}
	V(r)= -1 + \frac{\frac{E^2}{m^2} g_{\varphi\varphi} + 2g_{t\varphi}\frac{EL}{m^2} + \frac{L^2}{m^2} g_{tt} }{g^2_{t\varphi} - g_{tt}g_{\varphi\varphi}},
\end{equation}
and 
\begin{equation}
	{\hat g}_{rr}= g_{rr} + \frac{g_{tt}}{g^2_{t\varphi} - g_{tt}g_{\varphi\varphi}} g^2_{r\varphi}
\end{equation}

The angular velocity of the orbiting SCO  is defined as $\Omega\equiv u^\varphi/u^t=\dot{\varphi}$. For circular orbits $V(r)=0$ and $\frac{dV(r)}{dr}=0$. Then the orbital parameters $\{E,L,\Omega\}$ are given in a general form by
\begin{equation}
	\label{eqen}
	E=-m\frac{g_{tt}+g_{t\varphi}\Omega}{\sqrt{-g_{tt}-2g_{t\varphi}\Omega-g_{\varphi\varphi}\Omega^2}},
\end{equation}
\begin{equation}
	\label{eqan}
	L=m\frac{g_{t\varphi}+g_{\varphi\varphi}\Omega}{\sqrt{-g_{tt}-2g_{t\varphi}\Omega-g_{\varphi\varphi}\Omega^2}},
\end{equation}
\begin{equation}
	\label{eq:om}
	\Omega_\pm=\frac{-\partial_rg_{t\varphi}\pm\sqrt{(\partial_rg_{t\varphi})^2-\partial_rg_{tt}\partial_rg_{\varphi\varphi}}}{\partial_rg_{\varphi\varphi}},
\end{equation}
where the sign $\pm$ corresponds to prograde ($+$) and retrograde ($-$) orbits. 
 
The innermost circular orbit (ISCO) dwells at the smallest radius for which the orbit is marginally stable, meaning $\partial^2_rV=0$. Thus, it is found for the smallest radius $r=r_{ISCO}$ that satisfies the following equality,
\begin{equation}
	\label{vrr}
	\frac{E^2}{m^2}\partial_r^2g_{\varphi\varphi}+2\frac{EL}{m^2}\partial^2_rg_{t\varphi}+ \frac{L^2}{m^2}\partial_r^2g_{tt}-\partial^2_r\left(g_{t\varphi}^2-g_{tt}g_{\varphi\varphi}\right)=0.
\end{equation}

\begin{figure}
    \includegraphics[width=0.8\textwidth]{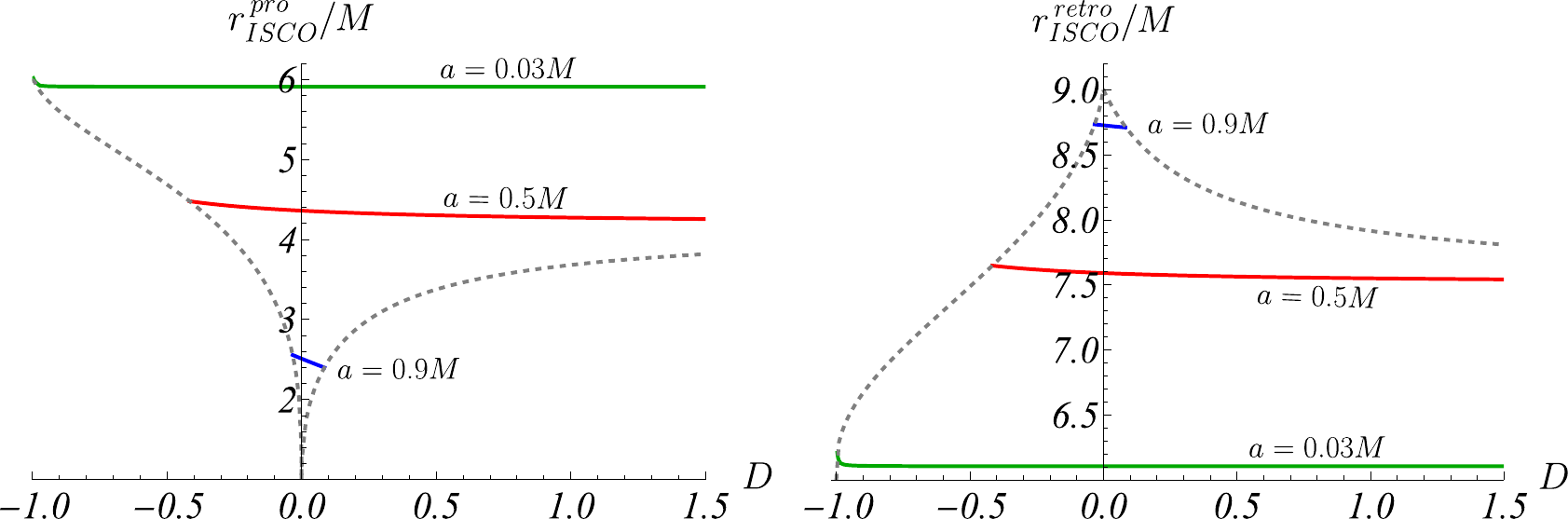}
	\caption{The radius $r_{ISCO}/M$ for prograde (\textit{left panel}) and retrograde orbits (\textit{right panel}) for different values of the disformal parameter $D$ and black hole spin parameter $a/M=0.03$, $a/M=0.5$ and $a/M=0.9$. The gray dashed lines correspond to those values of the disformal and spin parameters beyond which the black hole horizon doesn't exist.}
	\label{fig:ISCO_pro_retro_1} 
\end{figure}

In  Fig. \ref{fig:ISCO_pro_retro_1} we show the dependence of the ISCO radius on the disformal parameter $D$ for two different values of the BH spin parameter for prograde and retrograde orbits. As one can see, in both cases the radius of ISCO decreases with the increase of the disformal parameter.  

\subsection{Gravitational waves}

We shall study the emission of GW in the hybrid quadrupole approximation. In other words, we consider a SCO in circular motion around a SMBH losing energy through GW emission modeled by the well-known Einstein quadrupole formula. Therefore, this approximation takes into account the lowest order post-Newtonian dissipative effect and it has been widely used in the literature for both Kerr and beyond-Kerr spacetimes (see e.g. \cite{Glampedakis:2002cb,Glampedakis:2005cf,Sopuerta:2009iy,Pani:2011xj,Canizares:2012is,Chua:2017ujo,Collodel:2021jwi,Delgado:2023wnj,Destounis:2021mqv}).

We first consider both objects as point particles in a flat spacetime. The central SMBH with mass $M$ is fixed at the center of the coordinate system and the SCO with mass $m$ moves in circular orbits on the equatorial plane such that 
\be
x_1=x=R\cos\Omega t, \qquad x_2=y=R\sin\Omega t, \qquad x_3=z=0,
\ee
where $R$ is the radial coordinate and $\Omega$ is the angular velocity of the SCO. In this setup, the quadrupole moment tensor is given by \cite{Thorne:1980ru} 
\be
\label{eq:Istf}
\mathcal{I}_{ij}=\left[\int \rho x_ix_jd^3x\right]^{\textrm{STF}},
\ee
where STF means we cast the tensor in a symmetric trace-free form and $\rho=m\delta^3(\mathbf{x}-\mathbf{x}_\mathrm{SCO})$. The Einstein quadrupole formula for the GW energy flux is  given by,
\be
\label{eq:fen1}
\dot{E}=\frac{1}{5}\dddot{\mathcal{I}}_{ij}\dddot{\mathcal{I}}_{ij}.
\ee
For circular orbits, the energy lost due the GW emission is then 
\be
\label{eq:fen2}
\dot{E}=-\frac{32}{5}m^2R^4\Omega^6 = -\frac{32}{5}\mu^2 M^2R^4\Omega^6,
\ee
where $\mu$ is the mass ratio $\mu=m/M$.

The gravitational wave field, i.e. the metric perturbation $h^{TT}_{ij}$ in the  transverse traceless ($TT$) gauge is then given simply by 
\be
\label{eq:hijtt}
h^{TT}_{ij}=\frac{2}{{\cal D}_{L}}\ddot{\mathcal{I}}_{ij}^{TT},
\ee
where ${\cal D}_{L}$ is the luminosity distance to the observer and  the transverse traceless (TT) part of the tensor $h_{ij}$ is given by ,
\be
h_{ij}^{\textrm{TT}}=P_{ik}h_{kl}P_{lj}-\frac{1}{2}P_{ij}P_{kl}h_{kl},
\ee
where $P_{ij}=\delta_{ij}-n_in_j$ is the projection operator associated with the unit vector $n_i$  pointing from the observer to the source. 
The GW field $h^{TT}_{ij}$  can then be decomposed into plus and cross polarizations by defining the unit vectors $p$ and $q$ in the plane of the sky, namely

\be
\label{eq:uvectors}
p_i\equiv\frac{\epsilon_{ijk}n_jL_k}{|\epsilon_{ijk}n_jL_k|}, \qquad q_i\equiv\epsilon_{ijk}n_jp_k.
\ee
with $L_k$ being the unit vector parallel to the SCO angular momentum. The polarization tensors $H_{ij}^+$ and $H_{ij}^\times$ then read
\be
\label{eq:ptensors}
H_{ij}^+=p_ip_j-q_iq_j, \qquad H_{ij}^\times=p_iq_j+q_ip_j,
\ee

The GW field $h^{TT}_{ij}$ can be expanded into the polarization modes, such as
\be
\label{eq:hijtt+x}
h_{ij}^{\textrm{TT}}=A_+(t)H_{ij}^+ +A_\times(t)H_{ij}^\times,
\ee
and the wave's amplitudes can be expressed by
\be
\label{eq:amplitudes_general}
A_+(t)= - \frac{1}{2}H_{ij}^+h_{ij}^{\textrm{TT}}, \qquad A_\times(t)=\frac{1}{2}H_{ij}^\times h_{ij}^{\textrm{TT}},
\ee
For circular orbits on the equatorial plane, up to an unimportant constant phase, they become
\begin{eqnarray}
\label{eq:a+}
&&A_+=  \frac{4\mu M}{{\cal D}_{L}}  \Omega^2R^2   \left( \frac{1 + \cos^2\Theta}{2}\right) \cos\left(\Omega_{GW}t\right),\\
\label{eq:ax}
&&A_\times= \frac{4\mu M}{{\cal D}_{L}}  \Omega^2R^2 \cos\Theta \sin\left(\Omega_{GW} t\right),
\end{eqnarray}
where $\Omega_{GW}=2\Omega$ is the GW frequency and $\Theta$ is the inclination of the equatorial plane with respect to the observer. 

As a next step, we have to take into account the effect of the GW backreaction. The orbital radius and frequency are no longer constant because GW radiation energy is taken from the orbital energy of the SCO. We assume that the orbital evolution is adiabatic following a sequence of quasi-circular orbits until the plunge. Then the variation of the orbital energy as a variation of the orbital radius is easy to be found, namely
\begin{equation}\label{Edot}
	{\dot E}= \frac{\partial E}{\partial R} {\dot R}.
\end{equation}
With the help of the quadrupole formula (\ref{eq:fen1}), we can write the energy balance equation  as an equation describing the shrinking of the SCO orbit
\begin{equation}\label{Rdot}
	 {\dot R} = - \frac{32}{5}\frac{\mu^2 M^2R^4\Omega^6}{\frac{\partial E}{\partial R}},
\end{equation}
where $E$ and $\Omega$ are functions of $R$. Assuming that the shrinking of the  orbit is adiabatic,  the variation of the orbital angular velocity in time  is just $\Omega(t)=\Omega(R(t))$ where 
$R(t)$ is a solution to (\ref{Rdot}). The gravitational radiation backreaction changes also the GW signal, namely   \cite{Maggiore:2007ulw}

\begin{eqnarray}
	\label{eq:a+t}
	&&A_+=  \frac{4\mu M}{{\cal D}_{L}}  \Omega^2(t)R^2(t)   \left( \frac{1 + \cos^2\Theta}{2}\right) \cos\left(\Phi_{GW}(t)\right),\\
	\label{eq:axt}
	&&A_\times= \frac{4\mu M}{{\cal D}_{L}}  \Omega^2(t)R^2(t) \cos\Theta \sin\left(\Phi_{GW}(t)\right),
\end{eqnarray} 
where 
\begin{equation}\label{GWPHASE}
	\Phi_{GW}(t) =  2 \int^t_{0} \Omega(t^\prime) dt^\prime
\end{equation}
is the GW phase.

The last step is to apply the above flat spacetime GW theory to the curved spacetime of the SMBH. This can be done by an appropriate identification of the radial coordinate $R$ with some geometrical radial coordinate from the curved SMBH metric. The natural choice is to identify $R$ with the circumferential radius $\sqrt{g_{\varphi\varphi}}|_{\theta=\frac{\pi}{2}}$, i.e. from now on we have  $R=\sqrt{g_{\varphi\varphi}}|_{\theta=\frac{\pi}{2}}$. The energy involved in eqs.(\ref{Edot}) and (\ref{Rdot}) is naturally identified with the orbital energy  (\ref{eqen}).  

It is worth mentioning that if we identify the flat space coordinate $R$ with the Boyer-Lindquist radial coordinate $r$ the numerical results we present  below are in practice the same with those for circumferential radius. 

\section{Numerical results}\label{NR}

In this section, we study the qualitative and quantitative influence of the disformal parameter $D$ on the EMRIs and the associated gravitational waveforms.  We shall keep our study as general as possible without specifying the mass $M$ of the SMBH and the ratio $\mu$ whenever possible.   Both prograde and retrograde orbits are investigated. Results only for the former, though, are presented here, while the retrograde orbits are given in the Appendix. 

We start by analysing the evolution of the circumferential orbital radius $R$, that can be found by solving numerically eq. (\ref{Rdot}), as well as the emitted GW frequency $\Omega_{GW}$. As an initial condition, we chose $R(0)=10M$. The evolution of the angular velocity $\Omega(t)$, or equivalently the evolution of the gravitational angular frequency $\Omega_{GW}(t)=2\Omega(t)$, is found by substituting $R(t)$  in the expression (\ref{eq:om}) for $\Omega(r(R))$. Since the disformal BHs differ from GR only in the presence of rotation, we shall focus mainly on two nonzero values of the angular momentum -- $a/M=0.5$ and $a/M=0.9$. As we have already commented, the disformal BH should not deviate too much from Kerr BH and therefore we will limit ourselves to moderate negative values of the disformal parameter. That is why we take as a minimum value $D=-0.8$. The maximal $D$ corresponds either to the limit where no disformal BH solution exists or to a value beyond which we observe saturation to the limiting solution (\ref{limit_df1}).

\begin{figure}
	\includegraphics[width=0.925\textwidth]{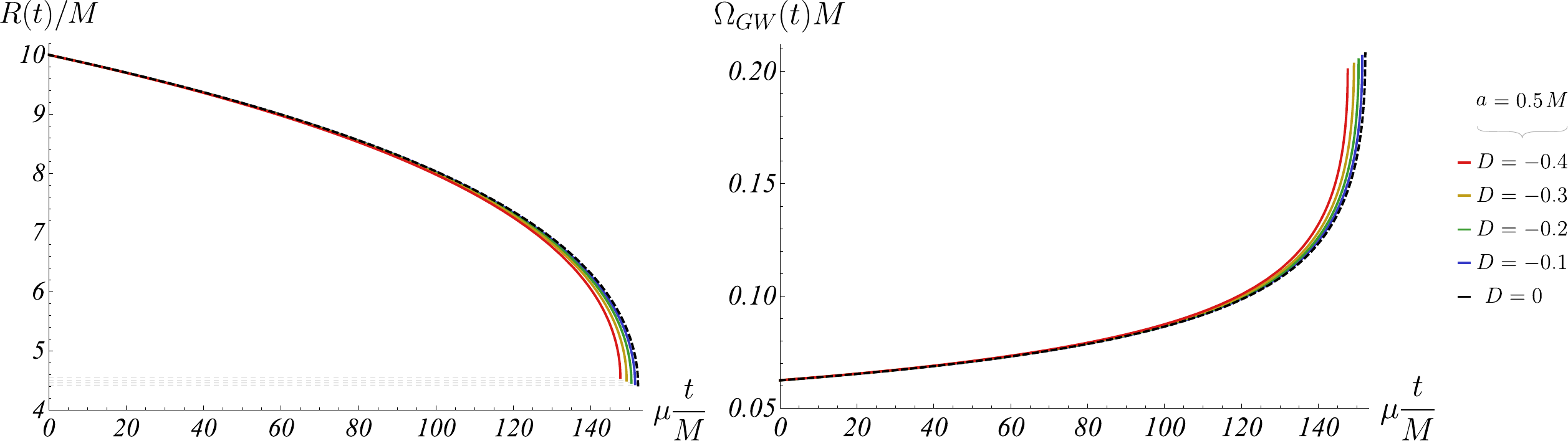}	
	\caption{Time evolution of the normalized circumferential radius  $R(t)/M$ and normalized GW angular frequency $\Omega_{GW}(t)M$ for different negative values of the disformal parameter $D$ and for  black hole spin parameter $a/M=0.5$ in the case of prograde orbits. The case of Kerr black hole is denoted by a dashed black line. }
	\label{fig:R_Om_a0.5_Dnegative}
\end{figure}

\begin{figure}
	\includegraphics[width=0.925\textwidth]{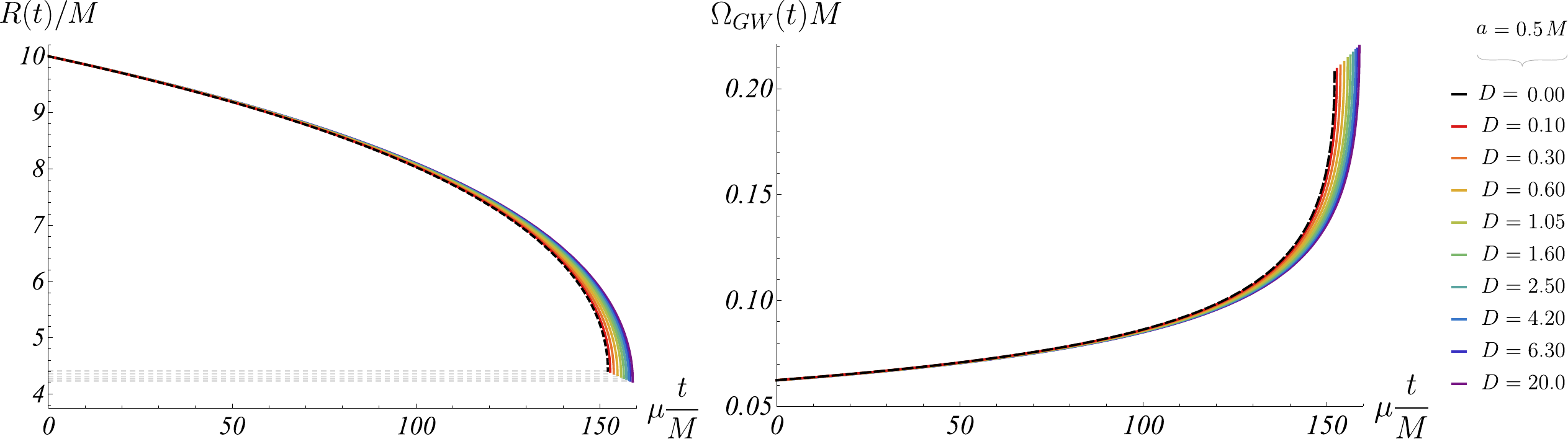}
	\caption{Time evolution of the normalized circumferential radius  $R(t)/M$ and normalized GW angular frequency $\Omega_{GW}(t)M$ for different positive values of the disformal parameter $D$ and for  black hole spin parameter $a/M=0.5$ in the case of prograde orbits. The case of Kerr BH is denoted by a dashed black line. } 
	\label{fig:R_Om_a0.5_Dpositive}
\end{figure}

In Fig. \ref{fig:R_Om_a0.5_Dnegative} we show the time evolution of $R$  and $\Omega_{GW}$   for different negative values of the disformal parameter $D$ and for black hole spin parameter $a/M=0.5$.  The case of Kerr black hole corresponds to $D=0$ and is denoted in the figure with a dashed black line. The inspiral begins at $R(0)=10M$ and then monotonically accelerates until it reaches the ISCO. Fig. \ref{fig:R_Om_a0.5_Dpositive} is the same as Fig. \ref{fig:R_Om_a0.5_Dnegative}, however for different positive values of the disformal parameter $D$. The largest value of $D$ presented on the figure is $D=20$ for which a strong saturation to the limiting solution (\ref{limit_df1}) is already observed.

\begin{figure}
	\includegraphics[width=0.925\textwidth]{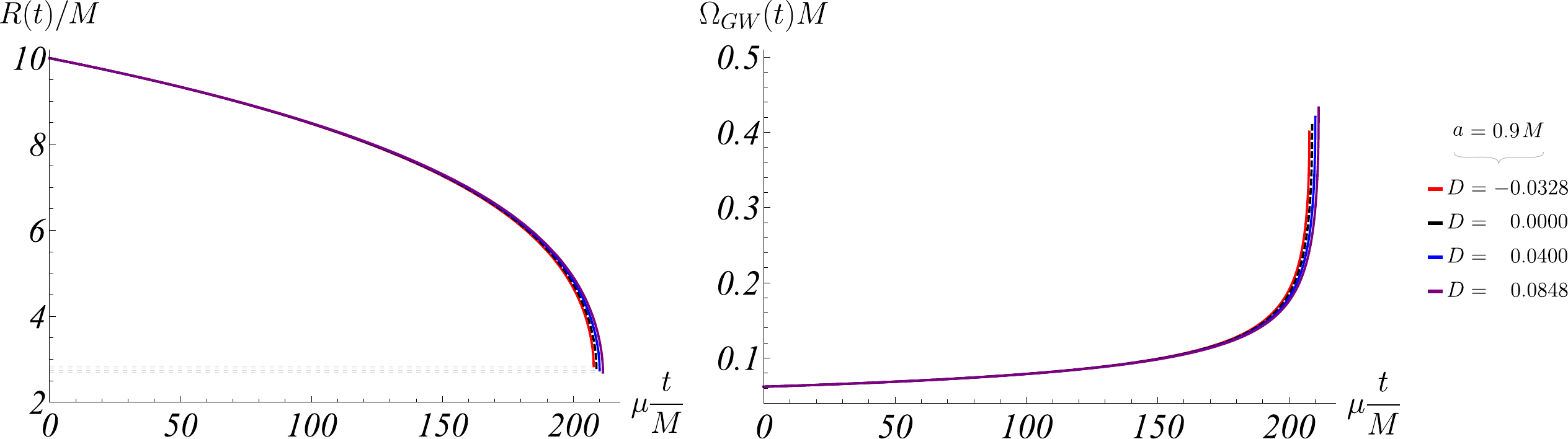}
	\caption{Time evolution of the normalized circumferential radius  $R(t)/M$ and normalized GW angular frequency $\Omega_{GW}(t)M$ for different negative and positive  values of the disformal parameter $D$ and for  black hole spin parameter $a/M=0.9$ in the case of prograde orbits. The case of Kerr BH is denoted by a dashed black line.} 
	\label{fig:R_Om_a0.9}
\end{figure}

In Fig. \ref{fig:R_Om_a0.9} we present the evolution of $R$  and $\Omega_{GW}$   for different, negative and positive, values of the disformal parameter $D$ and for black hole spin parameter $a/M=0.9$.  The Kerr black hole case is again denoted with a dashed black line. 
The largest positive value $D=0.3$ corresponds to the critical value of the disformal parameter beyond which no horizon exists for $a/M=0.9$. 

The discussed figures clearly show that the EMRIs can be  influenced by the disformal parameter $D$. Thus, the final values of $R$ and $\Omega_{GW}$ can differ  from those of Kerr BH.  The negative $D$ accelerates the inspiral in comparison with Kerr BH and the time of the plunge in this case is smaller compared to Kerr. The positive $D$, on the other hand, decelerates the inspiral and as a consequence, the time of the plunge is greater than that for Kerr BH. The leading reason for the described qualitative picture is the fact that the ISCO for the disformal BH is shifted compared to Kerr BH (see Fig. \ref{fig:ISCO_pro_retro_1}). In the considered interval for the disformal parameter, negative values of $D$ shift the ISCO towards larger radii compared to Kerr BH and vice versa for positive $D$.

\begin{figure}
	\includegraphics[width=0.8\textwidth]{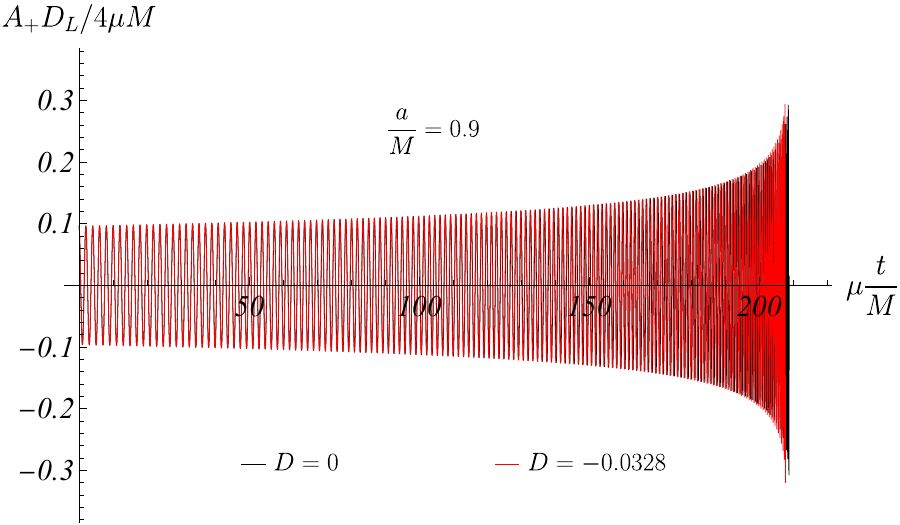}
        \includegraphics[width=0.8\textwidth]{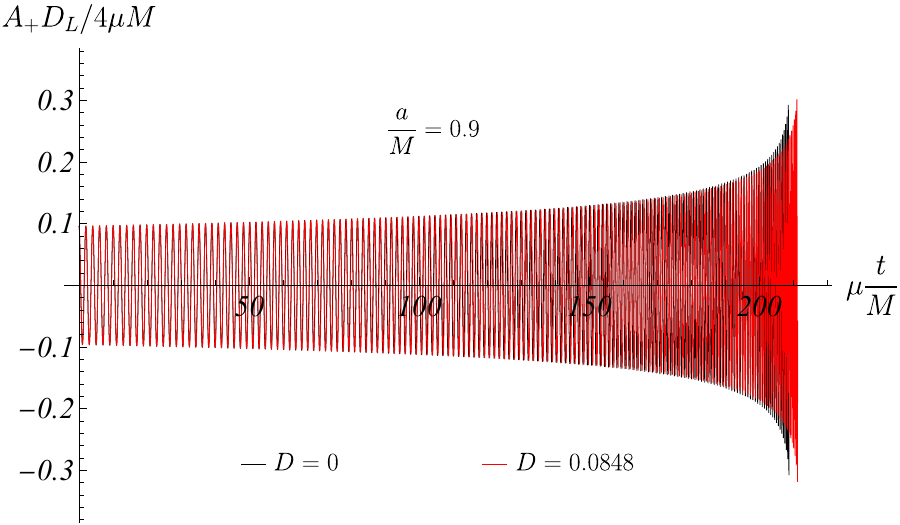}
	\caption{Waveforms for $\mu=2.5\times10^{-6}$ and $a/M=0.9$ in the case of prograde orbits and inclination angle $\Theta=0$. The black colour corresponds to the Kerr case while the red colour corresponds to the disformal Kerr with $D=-0.0328$ (top) and $D=0.0848$ (bottom). } 
	\label{fig:ApDL4muM_prograde_a0.9}
\end{figure}

The differences with the Kerr case are also visible in the waveforms that demonstrate the time evolution of both the GW phase and its amplitude. As an illustration,  in Fig. \ref{fig:ApDL4muM_prograde_a0.9}  we present the waveforms for two representative values of the disformal parameter $D$ while we keep $a/M=0.9$ with a fixed mass-ratio $\mu=2.5\times 10^{-6}$.  As seen in the figures, the effect on the GW amplitudes at the final stage, i.e. close to the ISCO, for negative values of $D$ is a bit smaller than the corresponding amplitudes for Kerr black hole while it is a  bit larger for positive $D$. 

In order to globally quantify the change of the gravitational waveforms caused by the presence of a disformal parameter with respect to the Kerr case, one can introduce a quantity $\Delta \Phi_{GW}$ called {\it dephasing} and defined as follows. Let $T_{ISCO}$ be the time interval  needed  for the SCO to reach the ISCO from 
its initial position $R(0)$. The accumulated gravitational phase during the whole inspiral from $R(0)$ is
\begin{equation}
	\Phi_{GW}= \int^{T_{ISCO}}_{0} \Omega_{GW}(t)dt .
\end{equation}
Since the evolution is adiabatic through sequence of quasi-circular orbits, $\Phi_{GW}$ is also given by 

\begin{equation}
	\Phi_{GW}= \int_{R(0)}^{R_{ISCO}} \Omega(R) \frac{dt}{dR} dR = \frac{5}{16} \int^{R_{0}}_{R_{ISCO}} \frac{\frac{\partial E}{\partial R}}{\mu^2 M^2 R^4 \Omega^5(R)} dR .
\end{equation}

The dephasing is then defined as
\begin{equation}
	\Delta \Phi_{GW}(D)=  \Phi^{disformal}_{GW}(D) -  \Phi^{Kerr}_{GW}.
\end{equation} 
Taking into account that $\Phi_{GW}/4\pi$ is the number $N$ of the cycles accumulated before the plunge, it is obvious that $\Delta \Phi_{GW}(D)/4\pi = N(D)-N^{Kerr}=\Delta N(D)$, i.e. the difference between the number of cycles made by the SCO around disformal Kerr BH and Kerr BH.  

\begin{figure}
	\includegraphics[width=0.9\textwidth]{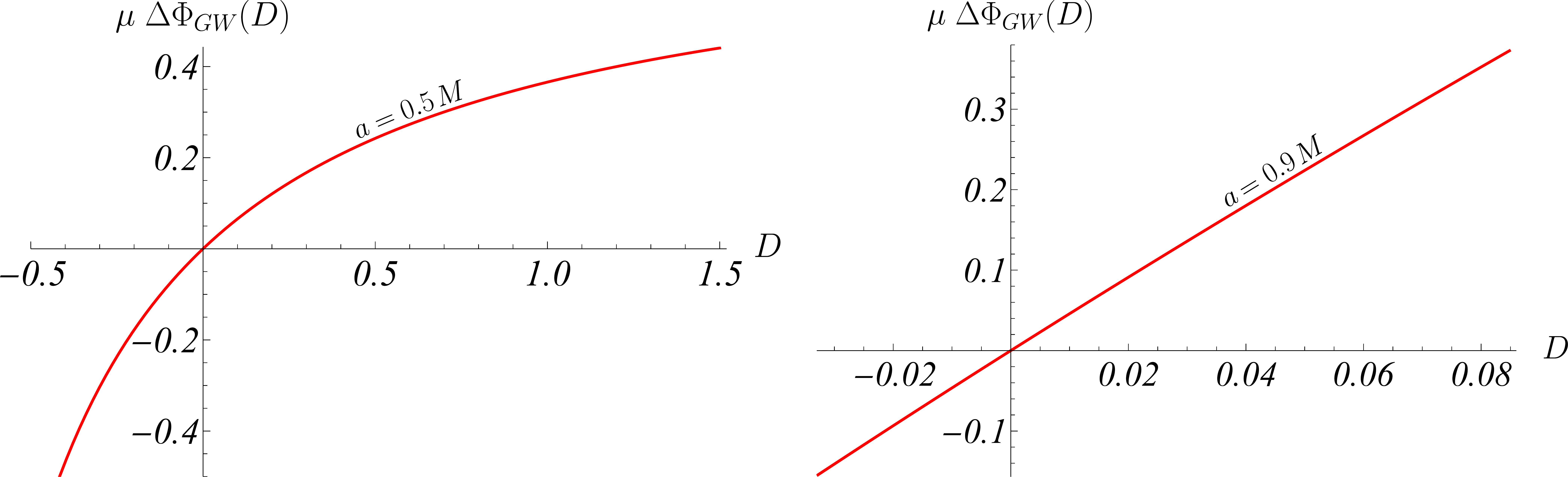}
	\caption{The dephasing for prograde orbits normalized to $\mu^{-1}$ as a function of the disformal parameter $D$. The left panel corresponds to $R(0)/M=10$ and $a/M=0.5$. The right panel is for  $R(0)/M=10$ and $a/M=0.9$. } 
	\label{fig:dephasing}
\end{figure}

Fig. \ref{fig:dephasing} shows the dependence of the dephasing $\Delta \Phi_{GW}(D)$ normalized to $\mu^{-1}$ on the disformal parameter $D$ for $R(0)=10M$ and two different spin parameters, namely $a/M=0.5$ and $a/M=0.9$.  In accordance with the other figures, the dephasing is negative for negative values of $D$ and positive for positive values of $D$. In other words, when $D$ is negative the number of cycles made by the SCO around the MBH is smaller than number of cycles round Kerr and vice versa for positive $D$.  

\section{Detecting  small disformal  deformations of Kerr black holes with LISA}\label{Detection}

Here we estimate the minimal disformal parameter $D$ that could  in principle be detected by LISA. In making such an estimate we assume that
the spacetime geometry deviates very little from the Kerr geometry  and the disformal parameter  $D$  satisfies $|D|\ll 1$. 

\begin{figure}
    \includegraphics[width=0.32\textwidth]{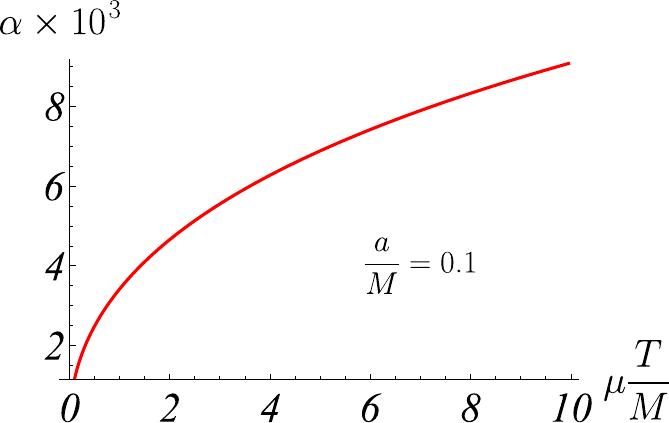}
    \includegraphics[width=0.32\textwidth]{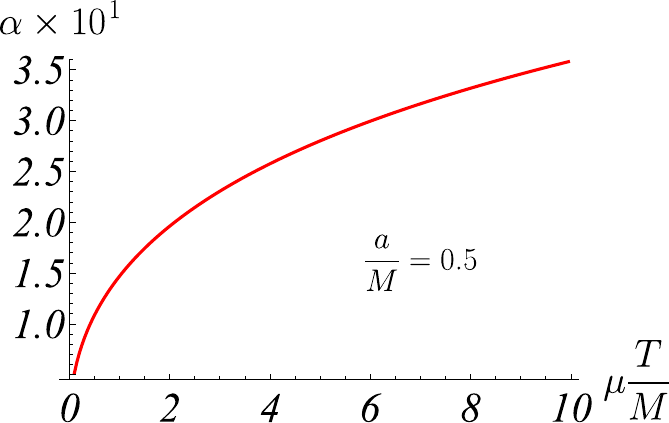}
    \includegraphics[width=0.32\textwidth]{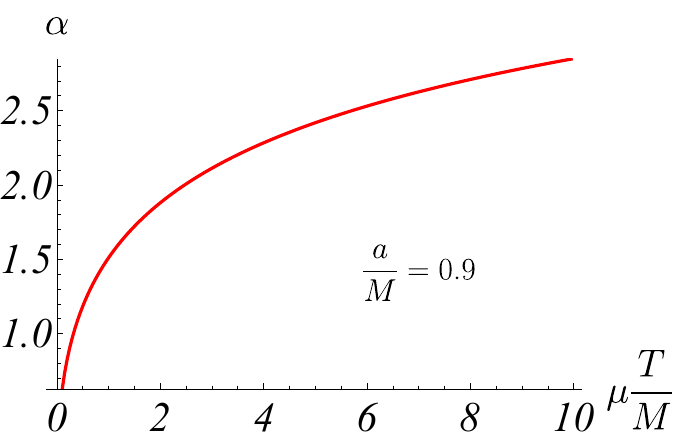}	
    \caption{\small The coefficient $\alpha$ for prograde orbits as a function of the ratio $\mu T/M$. The left panel corresponds to $a/M=0.1$, the center panel corresponds to $a/M=0.5$  while the right panel corresponds to $a/M=0.9$.}
	\label{fig:alpha_pro}
\end{figure}

\begin{figure}
    \includegraphics[width=0.32\textwidth]{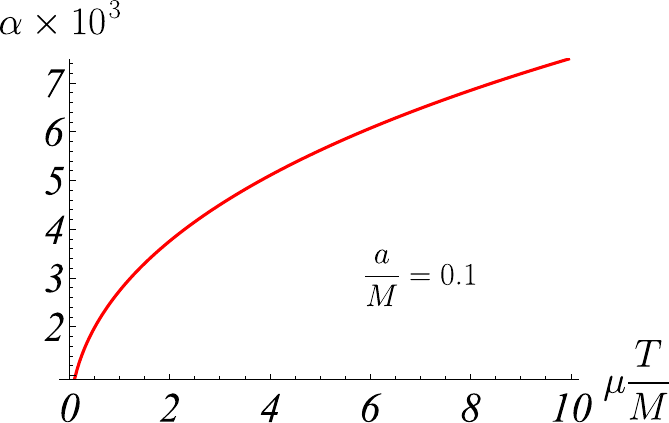}
    \includegraphics[width=0.32\textwidth]{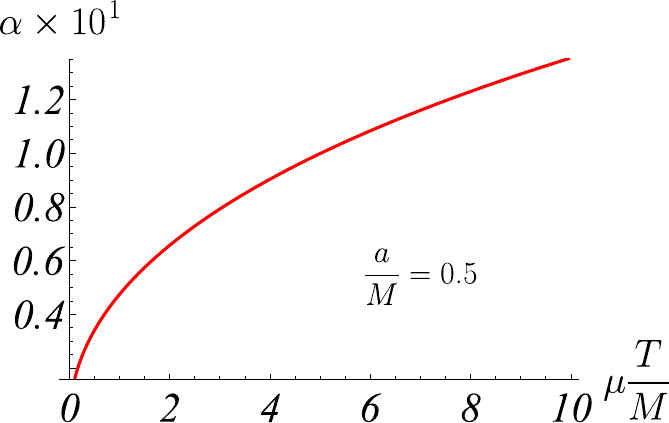}
    \includegraphics[width=0.32\textwidth]{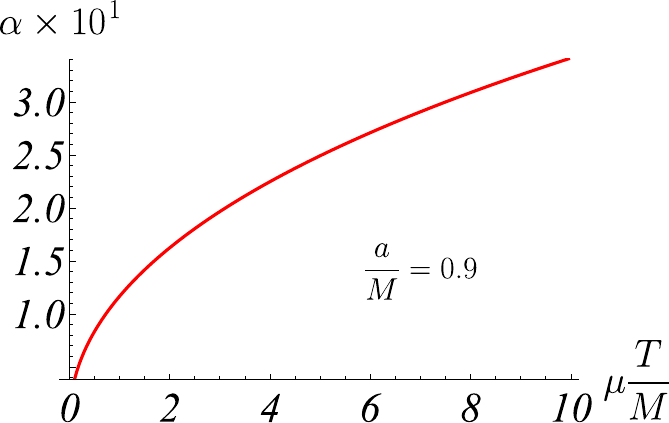}	
    \caption{\small The coefficient $\alpha$ for retrograde orbits as a function of the ratio $\mu T/M$. The left panel corresponds to $a/M=0.1$, the center panel corresponds to $a/M=0.5$ while the right panel corresponds to $a/M=0.9$.}
	\label{fig:alpha_retro}
\end{figure}

 Let us denote by $T$ the observational time in the LISA band before the plunge which should be smaller than the time of LISA mission operation, i.e. $T\le 4 {\rm yrs}$. Then expanding  $\Delta\Phi_{GW}(T,D, M, a)$ 
and retaining only terms linear in $D$ we find 
\begin{eqnarray}
	\mu \Delta\Phi_{GW}=\alpha D,
\end{eqnarray}
where the coefficient $\alpha$ depends on the ratios $\mu \frac{T}{M}$ and $\frac{a}{M}$, namely 
\begin{eqnarray}
	\alpha=\alpha\Big{(}\mu \frac{T}{M}, \frac{a}{M}\Big{)}.
\end{eqnarray}
The dependence of $\alpha$ on the ratio $\mu \frac{T}{M}$ for two different values of the spin parameter $\frac{a}{M}$ is presented in Fig. \ref{fig:alpha_pro} and Fig. \ref{fig:alpha_retro} in both cases of prograde and retrograde orbits.

The LISA sensitivity to the phase resolution of EMRI measurements is estimated as $\Delta\Phi_{GW}\sim 0.1 \, {\rm rad}$ assuming the signal to noise ratio (SNR) to be $SNR\ge 30$ \cite{Bonga:2019ycj}. Therefore the minimal disformal parameter $|D_{min}|$ that could in principle be detected by LISA is 
\begin{eqnarray}
	|D_{min}| \sim 0.1 \, \frac{\mu}{\alpha}. 
\end{eqnarray}

Let us consider some examples for 1-year observation in the LISA band. Natural candidates for the orbiting SCO are stellar  BHs and we choose $m=10 M_\odot$. For SMBH with mass of the order of $10^7 M_{\odot}$ and spin parameter between $a/M=0.5$ and $a/M=0.9$, our estimate shows that   minimal disformal parameter is of the order of  $10^{-7}$ for prograde orbits and $10^{-6}$ for retrograde orbits. 

The second example is for the SMBH SgrA* in the center of our galaxy with a mass  $M\approx 4.1\times 10^{6} M_\odot$. The spin of Sgr A* is not precisely known, but it is expected to be small, satisfying the upper bound $\frac{a}{M}\le 0.1$ \cite{Fragione:2020khu}. We shall consider the spin parameter in the interval $(0.05, 0.1)$. For these interval, the minimal disformal parameter that could be detected by LISA in the case of our galaxy, varies from  $10^{-5}$ to $10^{-4}$ for both prograde and retrograde orbits. 

\section{Discussion}

In this paper we have studied  EMRIs around disformal Kerr black holes within the framework of the quadrupole hybrid formalism. It was shown that the disformal deformation of Kerr black holes can strongly influence the EMRIs. Even when the disformal parameter is very small, its  effect on the globally accumulated phase of the gravitational waveform of the EMRIs can be significant and LISA will in principle be able to detect and measure very small values of the disformal parameter. i.e. very small deformations of the Kerr solution. Our estimation of the minimal disformal parameter that could be detected by LISA depends on the parameters of the SMBH  and can vary from $10^{-7}$  to $10^{-6}$ for SMBH with $M\sim 10^{7} M_\odot$ and spin parameter $0.5\ge \frac{a}{M}\le 0.9$. In the case of Sgr A* the minimal disformal parameter varies from $10^{-5}$ to  $10^{-4}$ for spin parameter $0.05\le \frac{a}{M}\le 0.1$. 

We have focused on the simplest case of SCO 's, those moving in the equatorial plane of a supermassive disformed Kerr black hole. We considered furthermore only zero eccentricity trajectories. One can except already that when eccentricity is present that the effects of the deformability parameter will be even greater as the local radial component $g_{rr}$ will also play a role (see for example \eqref{gnew2}). For equatorial motion, similarly to Kerr spacetime, timelike and null geodesics are separable  and spacetime is circular (as we discuss in the appendix). This is not the case for geodesic trajectories which are not equatorial for a disformal Kerr metric. Lightlike and timelike geodesics are not separable and therefore trajectories of SCO 's away from the equatorial plane are an open and uncharted territory. Similar and maybe not unrelated is the issue of non circularity of the disformal Kerr metric and what is the effect of this in the possible trajectories of EMRI systems. Work discussing the existence of photon non equatorial orbits which are circular or spheroidal  and their relation to separability of geodesics have been undertaken in the literature \cite{Glampedakis:2018blj}. In this direction one could enquire the existence of spherical or spheroidal trajectories for disformal Kerr metrics. What would be the effect of such orbits for EMRI 's? Such symmetric orbits have only been studied under the hypothesis of circularity. What would be the physical effects of non circularity for the existence and properties of such orbits?  It has been asserted in the literature that breaking of integrability and circularity leads to chaotic dynamics \cite{Chen:2023gwm}. It would be interesting to study these questions for the case of disformal Kerr metrics and we hope to do so in future work. 	
	
	\section*{Acknowledgements}
	This study is in part financed by the European Union-NextGenerationEU, through the National Recovery and Resilience Plan of the Republic of Bulgaria, project No. BG-RRP-2.004-0008-C01. DD  acknowledges financial support via an Emmy Noether Research Group funded by the German Research Foundation (DFG) under grant
	no. DO 1771/1-1. EB and CC acknowledge support from the french ANR grant StronG (ANR-22-CE31-0015-01).
\appendix 

\section{Retrograde orbits}

\begin{figure}
    \centering
    \includegraphics[width=0.925\linewidth]{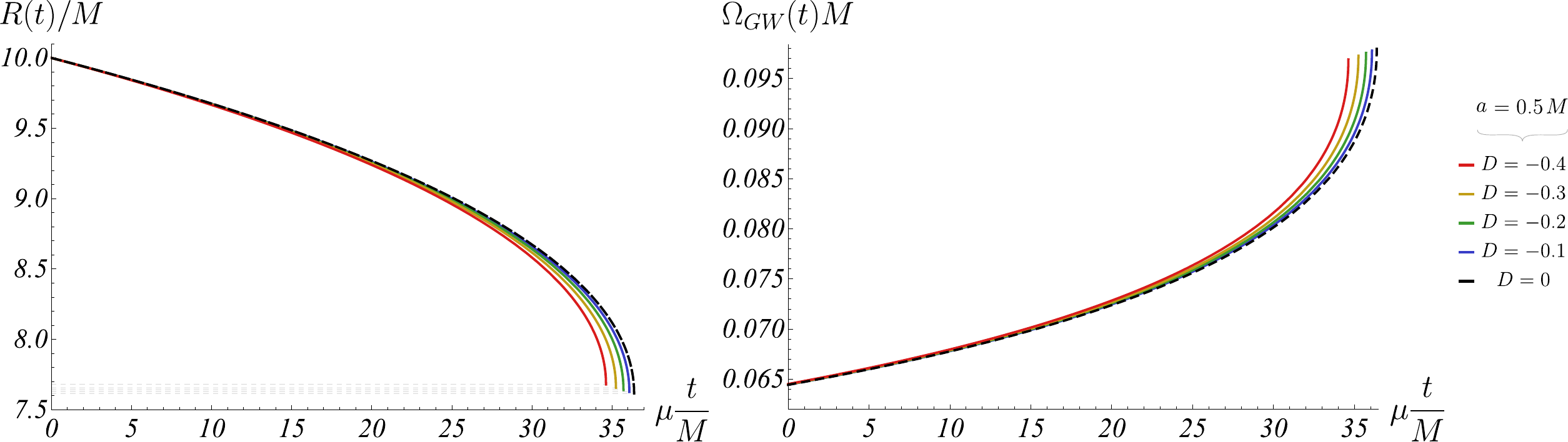}
    \caption{Time evolution of the normalized circumferential radius  $R(t)/M$ and normalized GW angular frequency $\Omega_{GW}(t)M$ for different negative values of the disformal parameter $D$ and for  black hole spin parameter $a/M=0.5$ in the case of retorgrade orbits. The case of Kerr BH is denoted by a dashed black line.}
    \label{fig:R_Om_a0.5_Dnegative_retro}
\end{figure}

\begin{figure}
        \centering
        \includegraphics[width=0.925\linewidth]{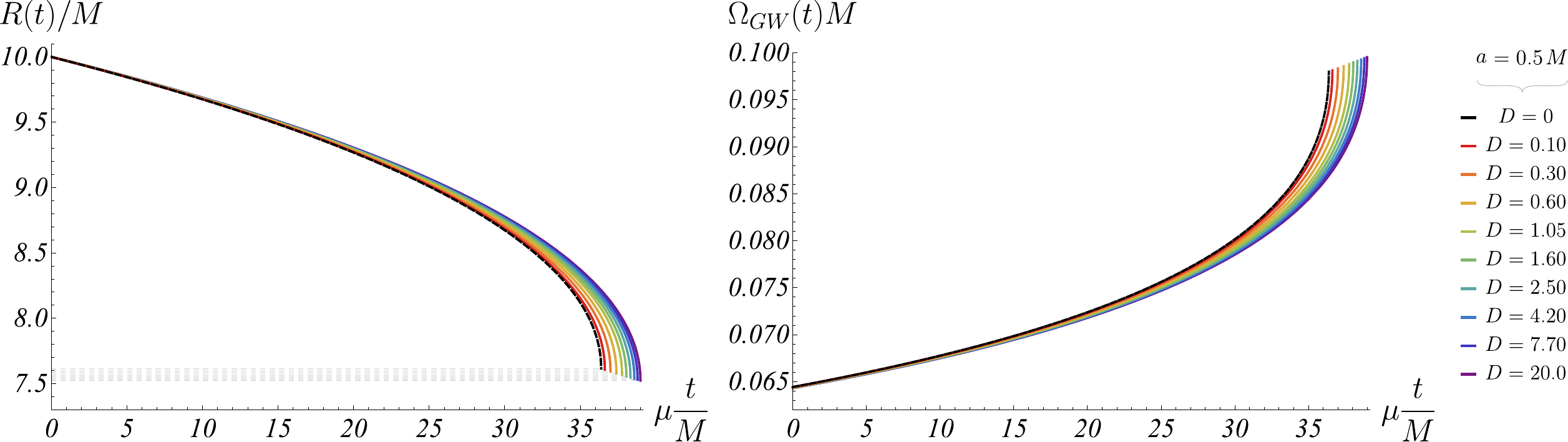}
        \caption{Time evolution of the normalized circumferential radius  $R(t)/M$ and normalized GW angular frequency $\Omega_{GW}(t)M$ for different positive values of the disformal parameter $D$ and for  black hole spin parameter $a/M=0.5$ in the case of retrograde orbits. The case of Kerr BH is denoted by a dashed black line. } 
	\label{fig:R_Om_a0.5_Dpositive_retro}
    \end{figure}

\begin{figure}
    \centering
    \includegraphics[width=0.925\linewidth]{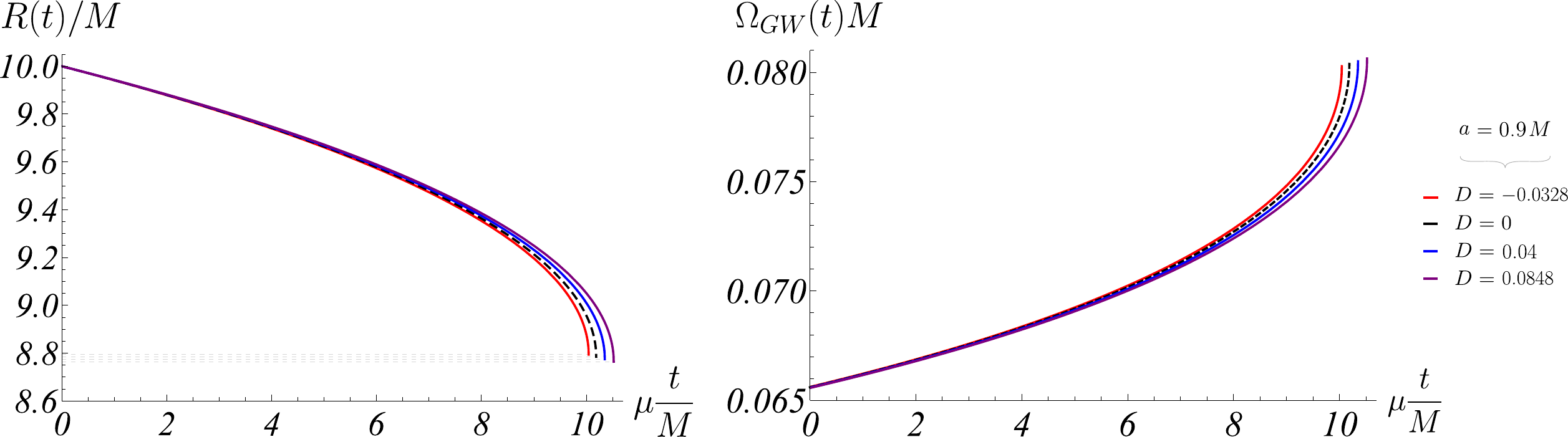}
    \caption{Time evolution of the normalized circumferential radius  $R(t)/M$ and normalized GW angular frequency $\Omega_{GW}(t)M$ for different positive values of the disformal parameter $D$ and for  black hole spin parameter $a/M=0.9$ in the case of retrograde orbits. The case of Kerr BH is denoted by a dashed black line.} 
	\label{fig:R_Om_a0.9_retro}
\end{figure}

\begin{figure}
    \centering
    \includegraphics[width=0.9\linewidth]{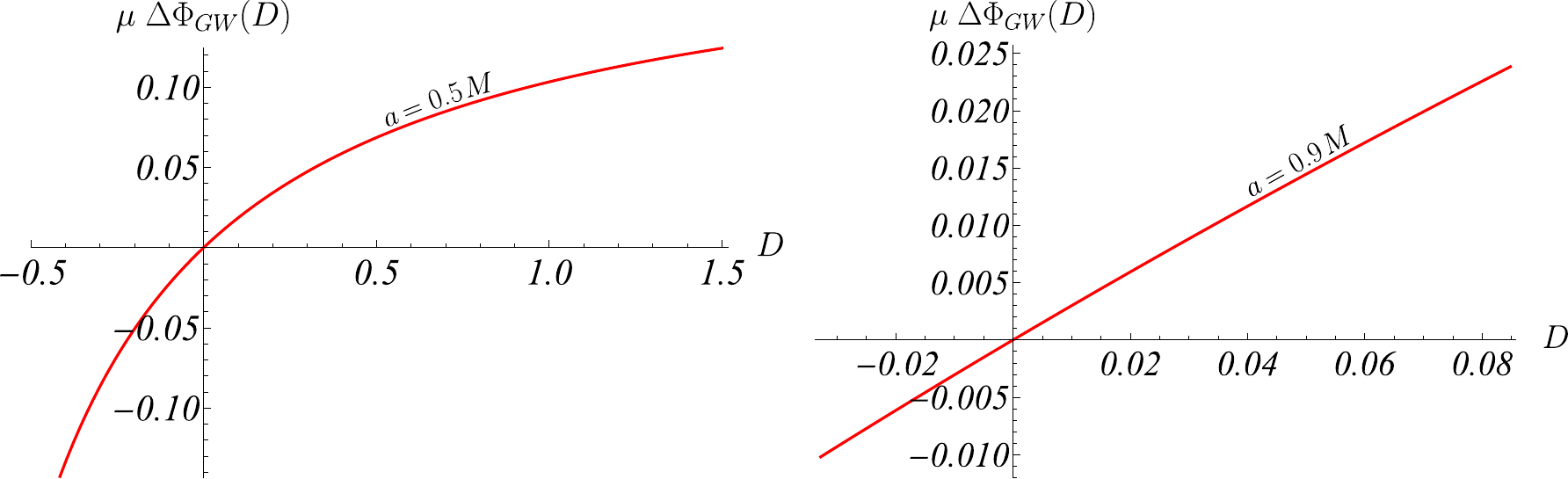}
    \caption{The dephasing for retrograde orbits normalized to $\mu^{-1}$ as a function of the disformal parameter $D$. The left panel corresponds to $R(0)/M=10$ and $a/M=0.5$. The right panel is for  $R(0)/M=10$ and $a/M=0.9$.} 
	\label{fig:dephasing_retro}
\end{figure}

Here we present the numerical results for the evolution of the circumferential radius and the GW frequency in the case of retrograde orbits. As one can see in Fig. \ref{fig:R_Om_a0.5_Dnegative_retro} - Fig. \ref{fig:R_Om_a0.9_retro}, the qualitative picture is very similar to that for the prograde orbits. As for prograde orbits, the negative values of the disformal parameter $D$ accelerate while the positive values of $D$ decelerate the inspiral in comparison with Kerr BH as one can see in the figures. In comparison with the prograde case, the time of the plunge is smaller as a consequence of the fact that the ISCO is shifted towards larger radius compared to the prograde case (see Fig. \ref{fig:ISCO_pro_retro_1}). This reflects itself also in the dephasing shown in Fig. \ref{fig:dephasing_retro}  --  the number of cycles is less than that for the prograde orbits.  

\section{A circular coordinate system for the disformal Kerr equator}

Given that at the equatorial plane $\theta=\frac{\pi}{2}$ we have a three dimensional subspace we can always make the metric circular. This will not be true in general and therefore we do not use this coordinate system in the body of the paper. 
Indeed consider the change of variables which when applied to Eq.~(\ref{df}) gives,
   \begin{equation*}
   		d t \to d t  -
   	D\frac{\sqrt{2Mr(r^2 + a^2_*)} \left(r^2+a^2_* + \frac{2M_*a^2_*}{r}
		\right)}{\Delta\left(\Delta+2DMr\left(1+\frac{a_*^2}{r^2}\right)\right)} d r;\quad 
  d\varphi\to d\varphi -\frac{2DMa\sqrt{2Mr(r^2 + a^2_*)}}{\Delta\, r\left(\Delta+2DMr\left(1+\frac{a_*^2}{r^2}\right)\right)}dr.
   \end{equation*}  
Metric~(\ref{df}) under this change takes a very simple (circular) form at $\theta=\pi/2$,
\begin{equation}
	\begin{split}
		\label{gnew}
		 g_{\mu\nu} d x^\mu d x^\nu &= -\left(1-\frac{2M}{r}\right) d t^2 -\frac{4Ma}{r} d t d\varphi + \left[r^2+a^2_* + \frac{2M_*a^2_*}{r}
		\right] d\varphi^2
		+ \frac{r^2}{\Delta_{eq}} d r^2 .
	\end{split}
\end{equation}
where
$$
\Delta_{eq}=\Delta+2DMr\left(1+\frac{a_*^2}{r^2}\right).
$$
Note a useful expression (analogous to the Kerr case):
\begin{equation*}
\Delta_{eq} = g_{t\phi}^2-  g_{tt}g_{\phi\phi}
\end{equation*}
In terms of the physical mass and rotation this gives,
\begin{eqnarray}
	\begin{split}
		\label{gnew2}
		 g_{\mu\nu} d x^\mu d x^\nu = -\left(1-\frac{2M}{r}\right) d t^2 -\frac{4Ma}{r} d t d\varphi &+ \left[r^2+\frac{a^2}{1+D} + \frac{2M a^2}{r}
		\right] d\varphi^2 \\
		&+ \frac{r^2}{r^2+\frac{a^2}{1+D}-2Mr+ \frac{2DM a^2}{r(1+D)}} d r^2 
	\end{split}
\end{eqnarray}
This agrees with the Kerr equator metric when we set $D=0$. 
Given the conserved quantities $E$ and $L$ and choosing an affine parameter for the equatorial geodesic trajectories we find in turn the parametric equations,
\begin{eqnarray}
    \dot\varphi &=& -\frac{Eg_{t\phi}+Lg_{tt}}{\Delta_{eq}} =\frac{2Ma E+(r-2M)L}{r \Delta_{eq}}\\
    \dot t  &=& \frac{Eg_{\phi\phi}+Lg_{t\phi}}{\Delta_{eq}} = \frac{r E(r^2+\frac{a^2}{1+D}+\frac{2 M a^2}{r})-2 M a L}{r \Delta_{eq}}\\
    \dot r &=& \pm\frac{\sqrt{E^2 g_{\phi\phi}+2ELg_{t\phi}+L^2 g_{tt}-\Delta_{eq}m^2}}{r}\nonumber\\
    &=&  \pm \frac{\sqrt{-L^2(r-2M)-4 a MEL+E^2r (r^2+\frac{2Ma^2}{r}+\frac{a^2}{1+D})-m^2 r\Delta_{eq}}}{r^{3/2}}\label{alt}
\end{eqnarray}
which have a very similar form to the case of Kerr equatorial trajectories which are given for $D=0$. 
One can check that \eqref{eq:rd} and \eqref{alt} are one and the same equation since the radial coordinate is the same. 

\newpage

\providecommand{\href}[2]{#2}\begingroup\raggedright\endgroup

\end{document}